\documentclass[twocolumn,9pt,amsfonts,longbibliography]{revtex4-1}
\usepackage{graphicx}
\usepackage{amsmath}
\usepackage{graphics}
\usepackage{appendix}
\usepackage{amssymb}
\usepackage{amsfonts,epsfig}
\usepackage{dsfont}
\usepackage{natbib}
\usepackage{hyperref}
\usepackage{enumitem}
\usepackage{physics}
\usepackage{color}
\usepackage[mathscr]{eucal}
\usepackage{ulem}
\usepackage{physics}
\usepackage{xcolor}
\usepackage{verbdef}
\usepackage[most]{tcolorbox}
\usepackage{mdframed}
\usepackage{multirow}
\usepackage{cancel}

\newcommand{\tim}{\tau}

\newcommand{\ir}{{R}}
\newcommand{\il}{{L}}

\newcommand{\dtn}{\epsilon}

\usepackage{color}
\definecolor{dgreen}{RGB}{0,90,0}

\begin{document}

\title{Adiabatic electron charge transfer between two quantum dots in presence of 1/f noise}
\author{Jan A. Krzywda}\email{krzywda@ifpan.edu.pl}
\affiliation{Institute of Physics, Polish Academy of Sciences, al.~Lotnik{\'o}w 32/46, PL 02-668 Warsaw, Poland}
\author{{\L}ukasz Cywi{\'n}ski}\email{lcyw@ifpan.edu.pl}
\affiliation{Institute of Physics, Polish Academy of Sciences, al.~Lotnik{\'o}w 32/46, PL 02-668 Warsaw, Poland}

\begin{abstract}
Controlled adiabatic transfer of a single electron through a chain of quantum dots has been recently achieved in GaAs and Si/SiGe based quantum dots, opening prospects for turning stationary spin qubits into mobile ones, and solving in this way the problem of long-distance communication between quantum registers in a scalable quantum computing architecture based on quantum dots. 
We consider theoretically the process of such an electron transfer between two tunnel-coupled quantum dots, focusing on control by slowly varying the detuning of energy levels in the dots. We take into account the fluctuations in detuning caused by $1/f$-type noise that is ubiquitous in semiconductor nanostructures, and analyze their influence on probability of successful transfer of an electron in a spin eigenstate. 
With numerical and analytical calculations we show that probability of electron not being transferred due to $1/f^\beta$ noise in detuning is $\propto \sigma^2 t^{\beta-1}/v$, where $\sigma$ characterizes the noise amplitude, $t$ is the interdot tunnel coupling, and $v$ is the detuning sweep rate. 
Interestingly, this means that the noise-induced errors in charge transfer are independent of $t$ for $1/f$ noise. 
For realistic parameters taken from experiments on silicon-based quantum dots, we obtain the minimal probability of charge transfer failure between a pair of dots is limited by $1/f$ noise in detuning to be the on order of $0.01$. This means that in order to reliably transfer charges across many quantum dots, charge noise in the devices should be further suppressed, or tunnel couplings should be increased, in order to allow for faster transfer (and less exposure to noise), while not triggering the deterministic Landau-Zener excitation.  
\end{abstract}

\date{\today}
		
\maketitle

\section{Introduction}
Spin qubit based on quantum dots (QDs) have achieved the level of single- and two-qubit gate performance \cite{Veldhorst_Nature15,Martins_PRL16,Zajac_Science18,Watson_Nature18,Yoneda_NN18,Yang_NE19,Huang_Nature19,Cerfontaine_arXiv19} that makes them viable building blocks of a scalable quantum computer. 
However, scaling up to large quantum circuits in architecture based on QDs will require mastering of long-distance quantum communication between registers of a few qubits \cite{Friesen_PRL07,Benito_PRB16,Vandersypen_NPJQI17,Veldhorst_NC17}. While applying multiple SWAP gates \cite{Loss_PRA98,Petta_Science05,Sigillito_arXiv19} to subsequent spin qubits in a chain of quantum dots is the most conceptually straightforward proposal, 
the two most recently successful avenues for achieving this goal are either coherently coupling stationary spin qubits to flying qubits, specifically to microwave photons \cite{Mi_Science17,Mi_Nature18,Samkharadze_Science18}, or simply making electron spin qubits mobile in a controlled way. The latter can be achieved in polar materials such as GaAs with surface acoustic waves \cite{Hermelin_Nature11,McNeil_Nature11,Bertrand_NN16} making a single electron travel for up to 100 $\mu$m \cite{McNeil_Nature11} distance, or by gate voltage controlled transfer of an electron along a chain of quantum dots. This method was shown to allow for coherent spin transfer across three and four quantum dots in GaAs \cite{Sanchez_PRL14,Baart_NN16,Fujita_NPJQI17,Flentje_NC17}, and for charge transfer across nine quantum dots in Si/SiGe heterostructure \cite{Mills_NC19}. We focus here on this long-distance electron transfer, since it  works also for nonpolar semiconductors, such as silicon. Electron spin localized in a silicon-based structure experiences much less nuclear noise than in GaAs, and creation of coherently controlled silicon-based quantum dot spin qubits has recently achieved a high degree of success both in {Si/SiO$_2$} \cite{Yang_NE19,Huang_Nature19} and Si/SiGe structures \cite{Kawakami_PNAS16,Watson_Nature18,Yoneda_NN18}. 

The process of electron transfer along a chain of quantum dots \cite{Mills_NC19} naturally decomposes into basic building blocks of charge transfer between pairs of neighboring dots. Each such transition can be considered separately, as a process, in which charge transfer is caused by a change of detuning of energy levels of electrons localized in the two dots. The basic physics is that of Landau-Zener problem \cite{Shevchenko_PR10}: detuning controls the alignment of two energy levels, while interdot tunneling couples them, and successful transfer of electron from one dot to another corresponds to adiabatic evolution of the system driven by a detuning sweep \cite{Zhao_SR18,Ban2019}. 
When one considers a system with only a single low-energy level (i.e.~when spin and valley splittings of the electron are very large, and we approximately deal with a ``spinless electron''), the only limitation for the charge transfer time is provided by the above requirement of approximate adiabaticity of Landau-Zener transition. However, in an often practically relevant situation of a few low-energy levels of the electron playing a role (i.e.~finite spin and/or valley splittings), the system goes through more than one energy level anticrossing during the charge transfer \cite{Zhao_arXiv18,Cota_JPCM18,Mi_PRB18,Shevchenko_PRB18}, and the conditions for high probability of dot-to-dot electron transfer become more involved.

However, the description of charge transfer problem becomes truly interesting, and experimentally realistic, after the effects of noise, inevitably present in a semiconductor nanostructure, are included. As we discuss below, fluctuations of electron spin splitting due to interaction with nuclei of the host material, are relevant only for very slow detuning sweeps, when even deterministic dynamics of a multi-level system leads to unwanted features in charge transfer probability. Furthermore, these effects can be essentially completely removed by isotopic purification of silicon \cite{Muhonen_NN14,Yoneda_NN18,Struck_arXiv19}. The main expected source of problems in the electron transfer process is thus charge noise, that has been widely recognized as an important driver of spin dephasing \cite{Kawakami_PNAS16,Struck_arXiv19} and relaxation \cite{Borjans_PRAPL19} for silicon-based spin qubits, and it is also known to affect the coherence of GaAs based spin qubits \cite{Dial_PRL13,Malinowski_PRL17}. 

In this article, we investigate the fidelity of adiabatic electron transfer between two neighboring 
quantum dots in presence of realistic $1/f^{\beta}$ noise in energy detuning between the two dots. Such coupling to charge noise has been recognized as the dominant one for electrons in coupled quantum dots \cite{Dial_PRL13,Martins_PRL16}, with fluctuations of tunnel coupling between the dots being widely assumed to be much weaker and thus less relevant (for discussion of possible exception to this point of view see \cite{Huang_NPJQI18}). Using the common terminology for Landau-Zener problem, we are dealing with influence of $1/f$-type {\it longitudinal} noise on transition probability. It should be noted that most of research on effects of noise on Landau-Zener problem was focused on {\it transverse} noise \cite{Kayanuma_PRB98,Malla_PRB17} having white \cite{Kayanuma_PRB98} or Lorentzian spectrum \cite{Malla_PRB17}, with longitudinal noise achieving much less attention \cite{Malla_PRB17}, and the case of  $1/f^{\beta}$ spectrum (highly relevant for charge noise in nanostructures used in solid state quantum information processing \cite{Paladino_RMP14,Szankowski_JPCM17}) even less. We focus our attention here mostly on silicon-based quantum dots (as coherence times in Si are longer than in GaAs, and silicon architectures have better prospects for scalability \cite{Lawrie_arXiv19}, once it becomes possible to create spin qubits with industrial Si technology), but the presented theory is applicable also to  GaAs-based spin qubits. We stress that we focus on transfer errors induced by charge noise, with less attention devoted to other sources of transfer imperfection. For a more detailed recent discussion of those see Refs.~\cite{Zhao_arXiv18,Zhao_SR18}.

We consider the electron being initialized in its lowest orbital energy state energy state in one of the dots, its spin being in an eigenstate of the Zeeman Hamiltonian. Note that for concreteness, and due to high level of development of quantum dot {\it spin} qubits, we focus here on electron spin as the relevant lowest-energy degree of freedom, but it should be noted that the theory presented below can be easily applied to the case of a valley qubit \cite{Culcer_PRL12,Boross_PRB16}. 
After delineating the range of sweep rates that allow for low-error transfer of electron (being in one of two spin states) in the noiseless case, we focus on influence that $1/f$ noise in detuning has on probability of successful transport of charge from one dot to another. 
Such a noise has a strong low-frequency component that is irrelevant for charge transfer probability, but its high-frequency tail leads to occurrence of event in which a noise-induced fluctuation in rate of change of detuning leads to excitation to the higher-energy state, and consequently to the failure of electron charge transfer. We parametrize the latter by probability $p_{\text{L}}$ of the electron staying behind in the left QD, when the intention was to transfer it to the left dot. 
Using analytical an numerical calculations we show that $p_{\text{L}}$  has a simple dependence on rms of the noise, $\sigma$, the detuning sweep velocity, $v$, the interdot tunnel coupling, $t$, and parameter $\beta$ characterizing the noise spectral density $S(\omega) \! \propto \! 1/\omega^\beta$. We have $p_{\text{L}} \propto \frac{\sigma^2}{v}t^{\beta-1}$, which means that  for $\beta\! =\! 1$, that is most often encountered in Si quantum dots \cite{Yoneda_NN18,Struck_arXiv19}, the noise-induced error in charge transfer procedure is to a very good accuracy {\it independent} of $t$. 

The paper is organized in the following way. In Sec.~\ref{sec:model} we describe our model of the double dot system and describe the basic physics of electron transfer via adiabatic sweep of detuning. In Section \ref{sec:nonoise} we discuss nonideal transfer of the higher-energy spin eigenstate caused by presence of finite spin-flip tunneling $t'$, and show how this effect is modified by quasi-static noise in spin splitting. Section \ref{sec:noise} contains the central results of the paper concerning the influence of $1/f$ noise on electron transfer probability. We describe there our analytical and numerical calculations, and present results for realistic silicon-based quantum dots.  

\section{The model} \label{sec:model}
We consider two lowest energy spin states, labeled by $s\! =\! \pm$ index, in each of two adjacent quantum dots: left (L) and right (R) one with dot-dependent Zeeman splittings, $\Delta_L$ and $\Delta_R$. We assume that interdot energy detuning $\epsilon(\tau)$ can be controlled by gate voltages. The Hamiltonian in this four-dimensional subspace, written in basis of diabatic states $\ket{L_-},\ket{L_+},\ket{R_-},\ket{R_+}$, is given as a sum of the following terms:
	\begin{align}
	H_\epsilon(\tau) &= \frac{1}{2}\left(\epsilon(\tau) \sum_{s=\pm} \ketbra{L_s} - \ketbra{R_s}\right) \nonumber \\
	H_t&=\frac{t}{2}\big(\ketbra{L_+}{R_+} + \ketbra{L_-}{R_-}\big) \nonumber \\&+ \frac{t'}{2}\big(\ketbra{L_+}{R_-} + \ketbra{R_+}{L_-} \big) + \text{h.c.} \nonumber \\
	H_{\Delta} &= \frac{\Delta_L}{2}\big(\ketbra{L_+}-\ketbra{L_-}\big) \nonumber \\& + \frac{\Delta_R}{2}\big(\ketbra{R_+}-\ketbra{R_-}\big) ,
	\end{align}
	where $t$ is the spin-conserving interdot tunneling, while $t'$ describes tunneling event accompanied by a spin flip, allowed by spin-orbit interaction. We expect $t' \! \ll \! t$, due to weakness of spin-orbit interaction in semiconductors from which gated quantum dots are made (this applies even more strongly to Si compared to GaAs). Let us stress that we are ignoring here the existence of the valley degree of freedom \cite{Zwanenburg_RMP13} in silicon. This is motivated partly by the desire to simplify the problem, so that the basic physics that we discuss here is not obscured, and partly by the fact that recent experiments on Si/SiGe dots show large valley splittings \cite{Hollman_arXiv19}. When valley splitting is larger than the Zeeman splitting, the complications due to valley-orbit mixing in Si can be ignored, and one can focus on the two lowest-energy states: the spin-split states of the lower energy valley. Let us however note, that in case in which the valley splitting is much smaller than the spin splitting, the theory discussed here applies to the transfer of the electron initialized in one of two lowest-energy valley states (each having the same spin projection on the spin quantization axis), and $t'$ corresponds then to intervalley tunneling. The key difference with respect to the case of two spin eigenstates is that $t'$ is then not necessarily much smaller than $t$, and the relation between them depends on the roughness of the quantum well interface \cite{Culcer_PRB10,Zhao_arXiv18}.

\begin{figure}[tb]
	\centering
	\includegraphics[width = \columnwidth]{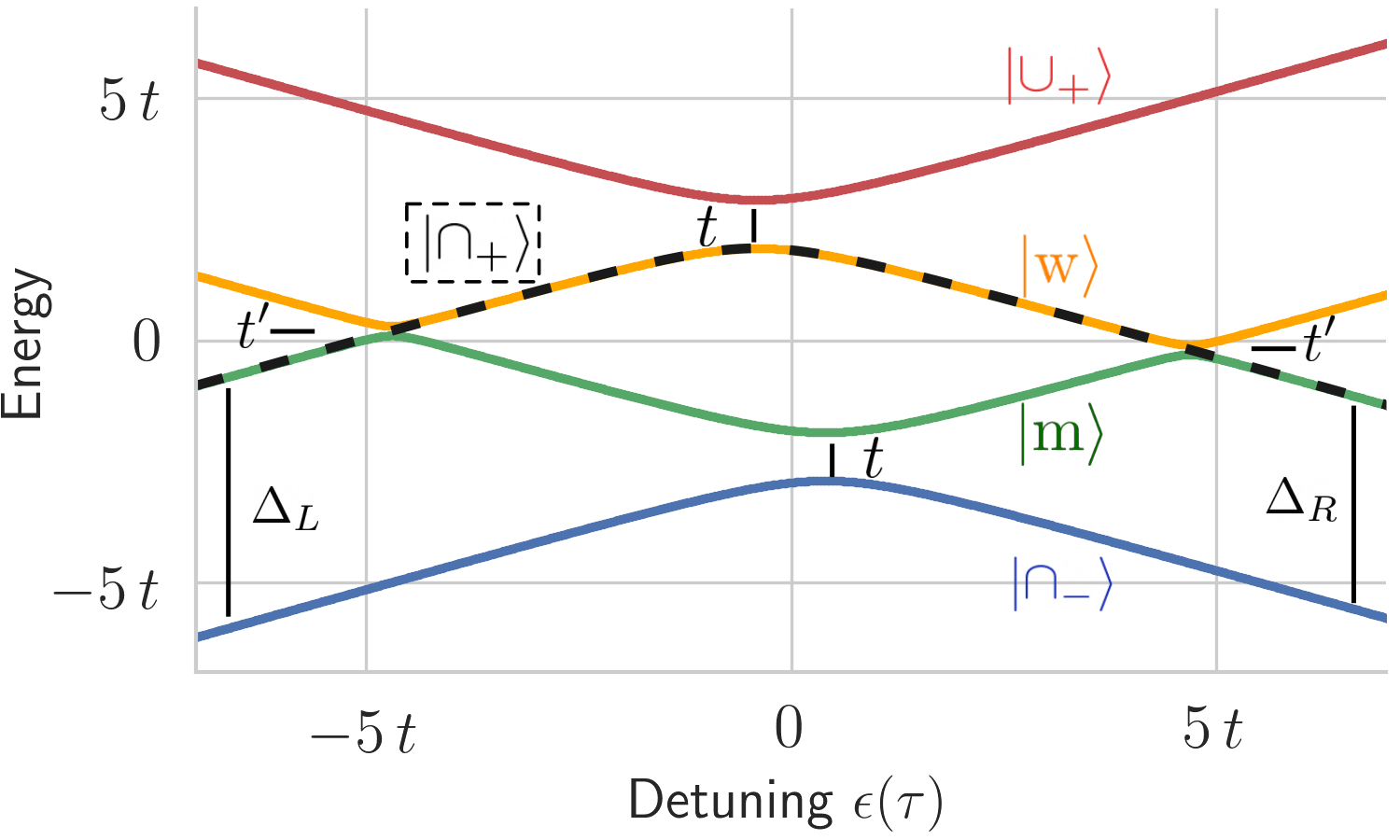}
	\caption{Schematic graph of detuning dependence of four lowest energy levels of a spin-split electron tunneling between the two quantum dots. On the far left side, the order of states, from the lowest-energy one, is $\ket{L_{-}}$, $\ket{L_{+}}$, $\ket{R_{-}}$, and $\ket{R_{+}}$, where $L/R$ denotes the state localized in left/right dot, and $s\! =\! \pm$ denotes the lower/higher energy spin state. 
	The tunnel coupling $t$ corresponds to spin-conserving transition between the dots, while $t' \! \ll \! t$ coupling is the spin-flip transition allowed by finite spin-orbit interaction. $\Delta_{L/R}$ are spin splittings in the left/right dot. Their difference, which leads to spin $+$ and $-$ anticrossing occurring at distinct values of $\epsilon$, is exaggerated compared to realistic situation in order to show its role. The colors and symbols ${\cap}_{-}$, $m$, $w$ and ${\cup}_{+}$ mark the energy levels that would be followed by a state initialized in one of eigenstates at negative $\epsilon$, if the change to positive $\epsilon$ was done very slowly, so that the electron is dragged adiabatically through all the level anticrossings. The ${\cap}_{+}$ path (dashed line) corresponds to evolution of the state initialized in the $\ket{L_+}$ state when detuning is changes at such a rate that the $t'$ transitions are diabatic, while $t$ anticrossing is traversed adiabatically. }
	\label{fig:states}
\end{figure}

Electron initialized in the left dot is now transferred to the right dot due to either spin preserving tunneling $t$ or tunneling with a spin-flip $t'$ combined with slow variation of detuning $\epsilon(\tau)$. We consider here the electron initialized in one of two $\ket{L_{\pm}}$ states. If the rate of change of $\epsilon(\tau)$ is slow enough, the adiabatic theorem applies, and $\ket{L_{-}}$ state should transform at long times into $\ket{R_{-}}$ state following the energy branch marked by ${\cap_-}$ in Fig.~\ref{fig:states}, while  $\ket{L_{+}}$ should transform into $\ket{R_{+}}$ following the energy branch marked there by $m$.

We assume that the time-dependence of $\epsilon(\tau)$ can be approximated by a linear one in the range of detuning shown in Fig.~\ref{fig:states}. We can use then the classical Landau-Zener result for probability of nonadiabatic excitation during the sweep of detuning through an anticrossing of energy levels. When such an anticrossing opens due to coupling of the two levels by $t$ matrix element ($t'$ in case of spin-flip transitions), the probability of such an excitation (equivalent to electron remaining in the left dot after the detuning sweep) is given by
\begin{equation}
p_{L} = \exp\left( - \pi \frac{t^2}{2 v}\right ) \,\, , \label{eq:PLZ}
\end{equation}
where $v$ is the sweep velocity, i.e.~$\epsilon(\tau) \! =\! \mathrm{const} + v\tau$ in the vicinity of the anticrossing, and we are using units in which $\hbar\! =\! 1$. Obtaining $p_{L} \! \ll \! 1$ for evolution of $\ket{L_{-}}$ state along the ${\cap}_{-}$ branch simply requires keeping $v$ below a certain value. For $m$ branch, on the other hand, we encounter three anticrossings, and it is the $t' \! \ll \! t$ splitting that limits the velocity $v$ if we want to follow this energy branch adiabatically. This might not be practical (especially in presence of noise, as we discuss in detail in this paper), and consequently the $\ket{L_{+}} \rightarrow \ket{R_{+}}$ transfer should rather be effected by a detuning sweep that leads to almost diabatic transition through $t'$ anticrossings (so that they effectively become level crossings), and to adiabatic transition through $\ket{L_{+}}$ - $\ket{R_{+}}$ anticrossing - in other words, the state should follow the path marked by ${\cap}_{+}$ in Fig.~\ref{fig:states}. We will return to this problem in the next Section, and now we will focus on the last, and most important, element of our model: the noisy character of detuning.

We set Zeeman splitting $\Delta \! \gg \! t$ large enough to consider two sets of anti-crossings between $\ket{L_-}$/$\ket{R_-}$ and  $\ket{L_+}$/$\ket{R_+}$ separately. 
Electron initialized in one of spin eigenstates is now transferred to the right dot due to spin-conserving tunneling term $t$ combined with slow variation of interdot energy detuning given by $\dtn(\tau) = E_{\il_-}(\tau) - E_{\ir_-}(\tau)$. This detuning is modified by the fluctuations of electric field $\dtn(\tau) = \dtn_0(\tau) + \xi(\tau)$, where $\xi(\tau) = \xi_L(\tau) - \xi_R(\tau)$. We stress here that although the fluctuations of the electric field in the vicinity of the dots lead to fluctuations of both $\dtn(\tau)$ and interdot tunneling, the latter is widely considered to be less important (see although Ref.~\cite{Huang_NPJQI18}).
Altogether the Hamiltonian reads
\begin{align}
\label{}
\hat H(\tim) &= \frac{\epsilon(\tau)+ \xi(\tau)}{2}\hat \sigma_z + \frac{t}{2}\hat  \sigma_x \equiv \frac{1}{2} \mathbf{r}_s(\tau) \cdot \vec{\sigma}
\end{align}
where we have introduced Pauli operators corresponding to dot degree of freedom, i.e. $\hat \sigma_z = \ketbra{L} - \ketbra{R}$. Evolution can be described in terms of effective field vector $\mathbf{r}(\tau)$, defined using its time-dependent magnitude and orientation angle $\theta(\tau)$:
\begin{align} 
|\mathbf{r}(\tau)| &= \sqrt{\big[\dtn(\tau) + \xi(\tau)\big]^2 + t^2 } \,\, , \label{eq:r} \\ 
\cot \theta(\tau) &=- \frac{\epsilon(\tau)  + \xi(\tau)}{t} \,\,  .  \label{eq:theta}
\end{align}
At any instant of time Hamiltonian $\hat H(\tau)$ can be diagonalized using time-dependent rotation $\hat R(\tim)\hat H(\tau)\hat R^{\dagger}(\tim) = \hat{\mathcal{H}}(\tau)$, which produces eigenenergies $E_{\pm}=\pm \tfrac{1}{2} |\mathbf{r}(\tau)|$ corresponding to the instantaneous eigenstates, denoted as ground state $\ket{g(\tau)} = \cos \tfrac{\theta(\tau)}2\ket{L} -\sin\tfrac{\theta(\tau)}2\ket{R}$, and excited state
$\ket{e(\tau)} = \sin \tfrac{\theta(\tau)}2\ket{L} +\cos\tfrac{\theta(\tau)}2\ket{R}$. Angle $\theta(\tau)$ allows to find the composition of the ground state, with probability of occupying left and right dot defined as $|\bra{L}\ket{g(\tau)}|^2 = \cos^2(\theta(\tau)/2)$ and $|\bra{R}\ket{g(\tau)}|^2 = \sin^2(\theta(\tau)/2)$ respectively. We consider now weak noise, specifically $\xi(\tau)/r_{0}(\tau) \! \ll \! 1$, where $r_{0}(\tau) =  \sqrt{\epsilon^2(\tau) +t^2} \! \geq \! t$ is the noiseless splitting between the adiabatic energy levels. Then we linearize  Eqs.~(\ref{eq:r}) and (\ref{eq:theta}) in $\xi/r_0$,
\begin{align}
\label{0thord}
|\mathbf{r}(\tau)| &\approx r_{0}(\tau) - \cos\vartheta(\tau)\,\xi(\tau)  \\\theta(\tau) &\approx \vartheta(\tau) + \sin \vartheta(\tau)\,\frac{\xi(\tau)}{r_0(\tau)}.
\end{align}
Noise independent angle $\vartheta(\tau) = \arccot(-\epsilon(\tau)/t)$, which defines relation between noiseless detuning and interdot tunneling, while $\vartheta(-\infty) = 0$ and  $\vartheta(\infty) = \pi$.
The time-dependent Hamiltonian, written using Pauli operators in the adiabatic basis, $\hat{\tau}_{z}(\tau) \! = \! \ket{e(\tau)}\bra{e(\tau)} -  \ket{g(\tau)}\bra{g(\tau)}$ etc., is given by
\begin{equation}
\hat{H}^{\text{ad}}(\tau) = \frac{r_{0}(\tau) + \xi_{\parallel}(\tau) }{2} \hat{\tau}_{z}(\tau) + \frac{\dot{\vartheta}(\tau) + \xi_{\perp}(\tau)}{2} \hat{\tau}_{y}(\tau) , \label{eq:Ha}
\end{equation} 
in which the longitudinal and transverse components of the noise are given by
\begin{align}
\xi_\parallel(\tau) & = \cos\vartheta(\tau) \xi(\tau)  \,\, , \label{eq:xilong} \\
\xi_\perp(\tau) & = \frac{\partial}{\partial \tau} \left[ \sin\vartheta(\tau) \frac{\xi(\tau)}{r(\tau)} \right] \,\, .  \label{eq:xiperp}
\end{align}	
Equation \eqref{eq:Ha} shows that transition between the state following an adiabatic trajectory from the initial one, to the other adiabatic state, can occur due to deterministic term $\propto \dot\vartheta \hat{\tau}_{y}$ (where $\dot \vartheta(\tau) \! =\! \text{d}\vartheta(\tau)/\text{d}\tau$), or due to noise-induced terms obtained by taking the derivative in Eq.~\eqref{eq:xiperp}. The calculation of transfer probability due to the deterministic term leads to the Landau-Zener formula, while the calculation of corrections to the transfer probability due to the noise-induced terms, which is the main topic of this paper, is considered in Section \ref{sec:noise}.

\section{Charge transfer in absence of fluctuations of detuning}  \label{sec:nonoise}
The probability of successful interdot transfer of an electron initialized in the lowest-energy state $\ket{L_{-}}$ is given in the absence of noise by Eq.~\eqref{eq:PLZ}. However, as we mentioned in the previous Section, even such a noiseless transfer becomes more interesting, if we consider an initial spin state of higher energy, $\ket{L_{+}}$.

\begin{figure}[tb]
	\includegraphics[width=\linewidth]{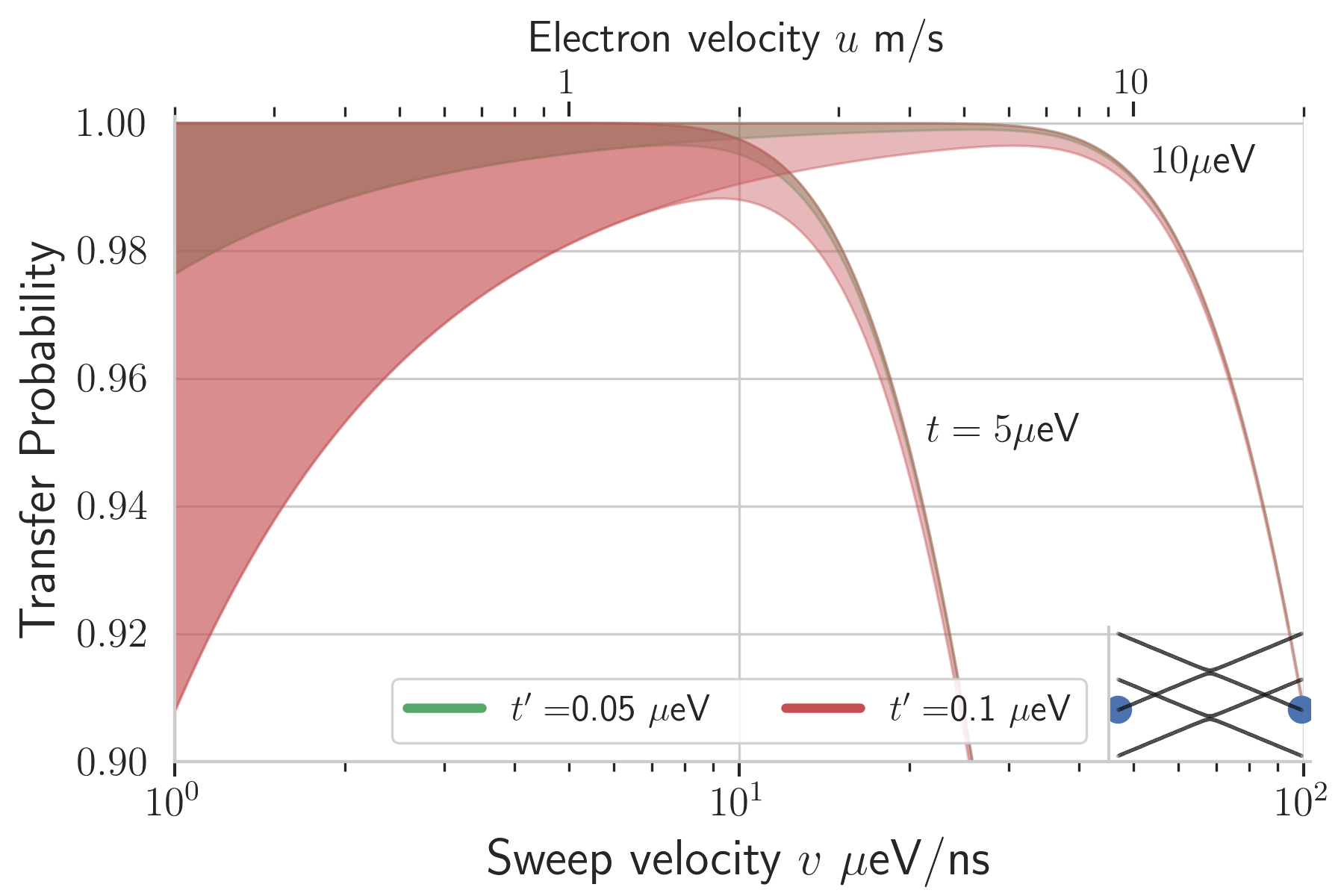}
	\caption{Transition probability in the noiseless case as a function of sweep rate for $t\! =\! 5$, $10$ $\mu$eV, $t' \! = \! 0.05$, $0.1$ $\mu$eV and $B\! =\! 1$ T. The oscillations of the probability are too fast to be discerned.
 }
	\label{fig:nonoise}
\end{figure}

In this case, there are two possible paths corresponding to successful transfer, marked in Fig.~\ref{fig:states} as $m$ and $\cap_+$, and probability is affected by their interference:
\begin{equation}
p(L_+\to R_+) = |a_\text m + a_{\cap_+}|^2 \,\, .
\end{equation}
In this equation $a_m$ is the probability amplitude of staying in the $m$-state for the whole experiment, while $a_{\cap_+}$ is the probability amplitude of temporal excitation to $w$-state between two minor anti-crossings $t'$. For sufficiently large $\Delta_{L/R}$ each transition can be regarded as independent thus the amplitudes read
\begin{align}
a_\text m &= \left(1 - p_{\text{LZ}}'\right) \sqrt{1 - p_{\text{LZ}} }\,e^{i\varphi_m} \nonumber \\
a_{\cap_+} &= p_{\text{LZ}}' \sqrt{1 - p_{\text{LZ}}}\, e^{i\varphi_m} e^{i \delta \varphi},
\end{align}
Where $p_{\text{LZ}}' = e^{-\pi t'^2/2v}$ and $p_{\text{LZ}}$ was defined in Eq.~\eqref{eq:PLZ}. Path difference is given by $\delta\phi \! =\! \int_{-\tau'}^{\tau'} E_{m}(\tau) - E_{\cap_+}(\tau)$ while $\tau'$ corresponds to a time at which first $t'$ anticrossing occurs.
For experimentally relevant parameters $\delta \varphi$ is expected to be very large, and can be estimated as the area of rhombus bounded by dashed $\cap_+$ and green $m$ line in Fig.~\ref{fig:states}, $\delta \varphi \sim \overline \Delta\times \overline \Delta / v =  \overline\Delta^2/v$. For illustration, if $\overline \Delta$ is caused by magnetic field of magnitude $1$ T at the sweep rate $v \approx 10 \mu$ev/ns, acquired phase is strongly oscillating and its value can be estimated as $\delta \sim  700\pi$. In particular if $\delta \varphi = \pi \mod (2\pi)$, destructive interference predicts lower bound of transfer probability
\begin{equation} 
p_{\min}(L_+ \to R_+) = \left(|a_{\text{m}}| - |a_{\cap_+}|\right)^2.
\end{equation}

These effects are illustrated in Fig.~\ref{fig:nonoise}, in which the transfer probability for the excited state is seen to oscillate very rapidly, with amplitude larger than $0.01$ for $v \! \approx \! 10$ $\mu$eV/ns. For value of $t' \! =\! 0.1$ recently reported for Si MOS quantum dots  \cite{Harvey-Collard_PRL19,Tanttu_PRX19}, the lower envelope of these oscillations reaches $0.9$ for $v \! \approx \! 1$ $\mu$eV/ns, showing that path-interference effects can have strong influence on transfer of higher-energy spin state in such dots.

	\begin{figure}
		\includegraphics[width = \columnwidth]{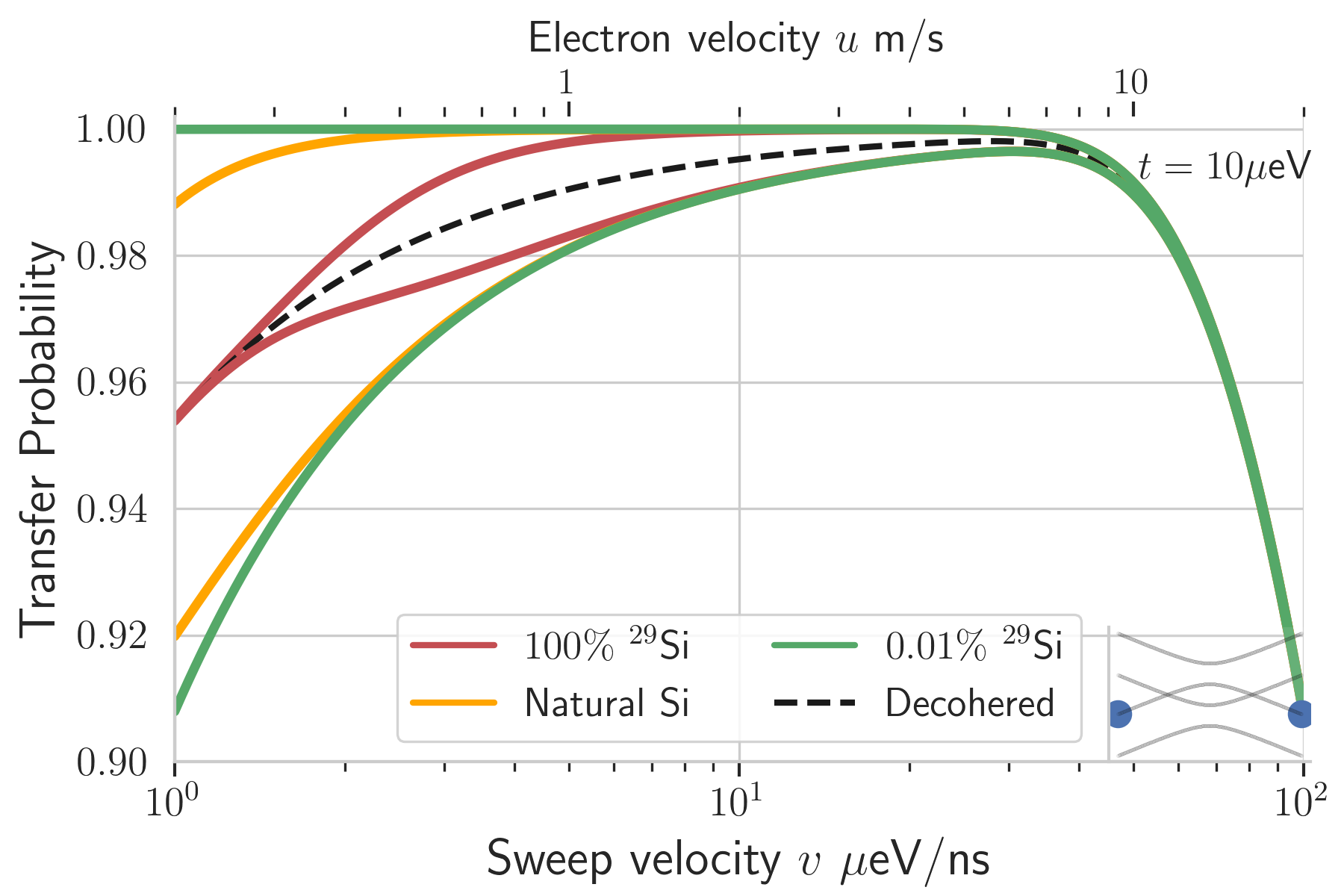}
		\caption{Envelope of transfer probability oscillations as a function of sweep velocity $v$ in a presence of quasistatic fluctuations of $\Delta_{L/R}$, for three different concentration of nuclear spins in silicon, with $\sqrt{\delta^2}$ from \cite{Assali_PRB11}, $t' = 0.1 \mu$eV and $B \!=\! 1$ T.}
		\label{fig:qsnoise}
	\end{figure}
	
In presence of phase noise, coherence between two paths can easily be lost, which eventually prevents any interference from happening. The result in this case is given by a classical sum of probabilities of traversing two paths:
\begin{equation}
\label{eq:cl}
p_{\text{cl}}(L_+ \to R_+)= |a_m|^2 + |a_{\cap_+}|^2 \,\, .  
\end{equation}	
Such dephasing can occur due to quasi-static fluctuations $\delta_{L/R}$ of spin splitting in each dot, resulting in $\Delta_{L/R} = \Delta_{L/R}^0 + \delta_{L/R}$. A common cause of such fluctuations is the presence of nuclear spins \cite{Hanson_RMP07,Cywinski_APPA11,Chekhovich_NM13}.
If we assume $\delta \varphi \sim \Delta_R \Delta_L/v$ and omit the term quadratic in noise (as $\Delta^{0} \gg \delta$), the noise correction yields $\delta \varphi' = (\Delta_L\delta_R + \Delta_R\delta_L)/v$. Now, since such fluctuations in each dot can be considered independent, quasistatic and Gaussian \cite{Merkulov_PRB02} with variance given by $\langle\delta_L^2\rangle = \langle\delta_R^2\rangle = \langle\delta^2\rangle $  (assuming equal size of the two dots), we can write for average transfer probability:
\begin{align}
	&\hspace{-0.3cm}\big\langle p(L_+ \to R_+)\big\rangle = |a_\text m|^2 + |a_{\cap_+}|^2 + 2 |a_\text m||a_{\cap_+}| \langle\cos(\delta \varphi)\rangle\, \nonumber \\=& p_\text{cl} + 2|a_m||a_{\cap_+}|\cos(\delta \varphi_0) \exp{-\sigma^2\frac{\Delta_L^2 + \Delta_R^2}{2v}}.
\end{align}
In Fig.\ref{fig:qsnoise}, we depict how the envelope of transfer probability oscillations is shrunk towards the classical result from Eq.~\eqref{eq:cl} as the rms $\sqrt{\langle \delta^2\rangle}$ of shifts of $\Delta_{L/R}$ is increased. The calculations are performed for silicon quantum dots with three different fractions of Si being the spinful $^{29}$Si isotope.

\section{Charge transfer in presence of dynamic noise} \label{sec:noise}
We focus now on the effect of fluctuations of detuning that occur on the timescale of electron transfer on adiabatic passage through the anticrossing between $\ket{L_s}$ - $\ket{R_s}$ states, characterized by $t$ coupling. As we could see in Eqs.~\eqref{eq:Ha} and \eqref{eq:xiperp}, nonzero {\it time derivative} of $\xi(\tau)$ couples the two adiabatic energy levels, and allows for finite probability of transition between them. 

\subsection{Model of $1/f$ charge noise}  \label{sec:1fmodel}
A commonly accepted microscopic model of $1/f^\beta$ charge noise in solid state nanostructures assumes that then noise is caused by multiple fluctuating dipoles, each of them treated as a two-level fluctuator (TLF): a system randomly switching with rate $\gamma$ between two states, each having a different dipole moment \cite{Schriefl_NJP06,Paladino_RMP14}. A microscopic nature of these TLFs in semiconductor devices hosting quantum dot spin qubits is not settled \cite{Beaudoin_PRB15}, but it is widely assumed that each such a TLF is a source of random telegraph noise (RTN), a non-Gaussian stochastic process characterized by correlation time $\tau_c$ and a Lorentzian power spectral density
\begin{equation}
S_{\text{TLF}}(\omega) = \frac{2\sigma^2_0 \gamma}{\gamma^2+\omega^2} \,\, , \label{eq:STLF}
\end{equation} 
where $\sigma_0$ is the rms of the noise, and $\gamma \! =\! 1/\tau_c$ is the switching rate of the TLF.
Assuming then that the probability of finding a fluctuator with given $\gamma$ is given by $N_{\beta}/\gamma^\beta$ in a wide range of frequencies, one obtains the $1/f^\beta$ power spectrum of noise caused by an ensemble of TLFs \cite{Schriefl_NJP06}. In this expression $N_{\beta}$ is a normalization constant, determined by equating $\int \text{d}\gamma N_\beta / \gamma^\beta $ to the total number of fluctuators, or alternatively (and more practically, when making contact with experiment), by adjusting the spectrum of noise to the observed one at a reference frequency. For the spectrum of noise generated by an ensemble of TLFs we obtain then
\begin{align}
S_{\beta}(\omega) & = \int_{\gamma_{\text{min}}}^{\gamma_{\text{max}}} \frac{N_{\beta}}{\gamma^\beta} \frac{2\sigma^2_0 \gamma}{\gamma^2+\omega^2} \text{d}\gamma \\ &= \frac{2\sigma^2_0 N_{\beta}}{\omega^\beta}  \int_{\gamma_{\text{min}}/\omega}^{\gamma_{\text{max}}/\omega} \frac{\text{d}x}{x^{\beta-1}(1+x^2)} \,\, ,
\label{eq:Sint} \\
&\approx g(\beta)\frac{2\sigma^2_0 N_\beta}{\omega^\beta}  \,\, ,  \label{eq:Sbeta}
\end{align}
where the dimensionless function $g(\beta)$ is given by the integral over $\text{d}x$ in Eq.~\eqref{eq:Sint}, with lower limit set to $0$, and upper limit set to $\infty$, which is applicable when the frequencies of interest fulfill $\gamma_{\text{min}}\! \ll \! \omega \! \ll \! \gamma_{\text{max}}$. It can be written as
\begin{equation}
g(\beta) \approx  \int_{0}^{\pi/2} \tan^{1-\beta}(\alpha) \, \text{d}\alpha \,\, ,
\end{equation}
from which it is easy to see that in  the range of $\beta$ of interest here this function takes values of the order of unity, as $g(1) \! =\! \pi/2$ and $g(3/2) = \pi/\sqrt{2}$.

For noise with $\beta\! =\! 1$, the $1/f$ character of spectrum typically extends to very low frequencies, and it is common to parametrize the noise by giving the value of $\sqrt{S(\omega_1)}\! \equiv \! \sqrt{S_1}$ at $\omega_1 \! =\! 2\pi$ $\frac{\text{rad}}{s}$ corresponding to 1 Hz. We have then $\sigma^2_0 N_1 \! =\! S_{1}\omega_1 /\pi$. On the other hand, for $1/f^\beta$ form of noise with $\beta\! > \! 1$ holding at rather high frequencies of interest ($\omega$ of the order of $t$, as we will see in the next Section), such a power-law behavior typically does not extend to very low frequencies. 
It is more convenient then to parametrize the noise by giving the value of $\sqrt{S(\omega_{h})}$ at some $\omega_h \! \gg \! \omega_1$ that is in the range of frequencies that have the strongest influence on the observable of interest. We have then $2\sigma^2_0 N_\beta = S(\omega_h)\omega^\beta_h / g(\beta)$.
 
The non-Gaussian nature of the RTN (i.e.~the fact that its higher-order correlation functions are not expressed by products of its autocorrelation functions, or equivalently that its power spectral density does not uniquely determine the properties of the noise \cite{Ramon_arXiv19}) becomes however relevant only when one considers strong coupling of TLFs to a system that is perturbed by them \cite{Galperin_PRL06,Ramon_PRB15,Szankowski_JPCM17}. Only then one has to go beyond the second order perturbation theory in calculation of influence of noise on the system of interest, and non-Gaussian features of RTN becomes apparent. While non-Gaussian features of $1/f^\beta$ charge noise affecting qubits have been a subject of theoretical attention \cite{Paladino_RMP14,Ramon_PRB15,Ramon_arXiv19}, their features observable in qubit coherence experiments are usually rather weak, with exception of the situations in which a single TLF is very strongly coupled to the qubit \cite{Nakamura_PRL02,Galperin_PRL06,Ramon_PRB12}. When this scenario is ruled out, it is reasonable to assume that the $1/f$ noise could be treated as Gaussian to a good approximation. In the following we make this assumption, and and by doing so we can replace RTNs by a Gaussian process, which share exactly the same power spectral density -  an Ornstein-Uhlenbeck process characterized by rms of $\sigma_0$ and correlation time $\tau_c$.

\subsection{Analytical approach to calculation of noise-induced transitions} \label{sec:analytical}
Let us consider the contribution of a single fluctuator, modeled as a source of Ornstein-Uhlenbeck noise $\xi_{\perp}$ characterized by $\sigma_0^2$ and $\tau_{c}\! =\! 1/\gamma$, to probability of transition to the other adiabatic energy level. We start by writing equations of motion for wavefunction of the form $\ket{\psi(\tau)} = c_+(\tau) e^{i\varphi(\tau)/2}\ket{+} \! +  c_-(\tau)e^{-i\varphi(\tau)/2} \ket{-}$, where dynamical phase is denoted as $\varphi(\tau) = \int_{-T}^{\tau} r_0(\tau) = \int_{-T}^{\tau} \sqrt{v^2 \tau^2 + t^2}$. The evolution is described by the adiabatic Hamiltonian given by Eq.~\eqref{eq:Ha}. We consider now the electron initialized in the lowest energy state (i.e.~$c_{-}(-\infty) \! = \! 1$, $c_{+}(-\infty) \! = \! 0$), but the calculation for the electron initialized in the excited state is analogous. 

For amplitude of electron being in excited adiabatic state, equation of motion reads:
	\begin{equation}
	\dot c_+(\tau) = i\,\frac{\xi_{\parallel}(\tau) c_+(\tau)}{2}  + \frac{\dot \vartheta(\tau) + \xi_{\perp}(\tau) }{2}  e^{-i\int_{-\infty}^{\tau} r_0(\tau)} c_-(\tau) \,\, .
	\end{equation}
	We seek now corrections to the adiabatic evolution, for which the influence of $\dot \vartheta(\tau)$ is by definition negligible. We thus omit $\dot\vartheta(\tau)$, and look for the solution in a perturbative way, by writing $c_\pm(\tau) = c^0_\pm(\tau) + \lambda c^1_\pm(\tau)+  \lambda^2 c^2_\pm(\tau) + \ldots$, where $\lambda$ counts the powers of noise, i.e.~we replace $\xi_{\perp/\parallel}(\tau)$ by $\lambda \xi_{\perp/\parallel}(\tau)$. In the zeroth order we have a perfectly adiabatic solution, for which $c^{(0)}_+(\tau) = 0$ and $c^{(0)}_-(\tau) = 1$, while the first order correction is given by 
	$$c_+^{(1)}(\infty) = \frac{1}{2} \int_{-\infty}^{\infty} \text{d}\tau\, \xi_{\perp}(\tau) e^{-i\int_{-\infty}^{\tau} r_0(\tau') \text{d}\tau'} \,\, .$$
From this result we obtain the  result for  $|c^{(1)}_{+}(\infty)|^2$ in the lowest order in $\xi_{\perp}$, and then we average it over realizations of noise, with averaging denoted by $\langle \ldots \rangle$: 
\begin{align}
\label{eq:first_ord}
\langle |c_+^{(1)}&(\infty)|^2 \rangle = \frac{1}{4} \int_{-\infty}^{\infty}\text{d} \tau_1  \text{d}\tau_2 \langle \xi_{\perp}(\tau_1)\xi_{\perp}(\tau_2) \rangle e^{-i\int_{\tau_2}^{\tau_1} \,r_0(\tau')} \nonumber\\
& \approx  \frac{t^2}{4} \int_{-\infty}^{\infty}\text{d} \tau_1  \text{d}\tau_2 \frac{\langle \dot \xi(\tau_1) \dot \xi(\tau_2) \rangle }{[r_0(\tau_1)r_0(\tau_2)]^2}e^{-i\int_{\tau_2}^{\tau_1} \text{d}\tau' \,r_0(\tau')} 
	\end{align}
In the second part we kept purely dynamical term proportional to noise derivative $\xi_\perp(\tau) \sim \frac{\sin\vartheta}{r} \dot \xi(\tau)$ and neglected possibly static contributions $\propto \xi(\tau)$, since slow noise is not expected to cause any excitation. The correlator of derivatives of noise is calculated in Appendix \ref{app:OUderivative}, while the integral in Eq.~\eqref{eq:first_ord} is done in Appendix \ref{app:OUcorrection}. The final result (which was also obtained using a different method in \cite{Malla_PRB17}) reads
\begin{equation}
|c^{(1)}_{+}(\infty)|^2 \approx \frac{\sigma^2_0}{v} R(t\tau_c) \,\, ,  \label{eq:cOU}
\end{equation}
where the function
\begin{align}
R(t\tau_c) 
= \frac{\pi}{2}t\tau_c\left(1 - \frac{1}{\sqrt{1+\frac{1}{t^2\tau_c^2}}}\right)
\end{align}
behaves as $R(t\tau_c) \! \approx \! \frac{\pi}{2} t\tau_c$ for $t\tau_c \! \ll \! 1$, while for  $t\tau_c \! \gg \! 1$ we have $R(t\tau_c) \! \approx \! \tfrac{\pi}{4} t\tau_c\,$ , and $R(t\tau_c)$ has  a maximum for $\tau_c \! = \frac{\pi}{2}\,0.78 /t   \! \approx 1/t$ . We see then that fluctuators having their switching rate $\gamma \! =\! 1/\tau_c$ close to the tunneling energy $t$ could be expected to have the strongest influence on nonzero value of $|c_{+}(\infty)|^2$, but for power-law noise spectral density, the cumulative contribution of fluctuators with much lower and higher switching rates is also significant. 

We use now the model of $1/f^\beta$ noise from Sec.~\ref{sec:1fmodel} to calculate $|c^{(1)}_{+}(\infty)|^2$ due to many fluctuators. Assuming that the influence of each fluctuator is so small that $\sum_{k}  |c_{+}(\infty)|_k^2 \! \ll \! 1$, we can still use the second order perturbation expansion in coupling to noise, and 
\begin{align}
\langle |c_{+}^{(1)}(\infty)|^2 \rangle_\beta & \approx \frac{\sigma^2_0}{ v} \int_{\gamma_{\text{min}}}^{\gamma_{\text{max}}} \frac{N_{\beta}}{\gamma^\beta} R(t/\gamma)  \,\, , \nonumber\\
& = \frac{t^2}{2 v} \int_{t}^{\infty} \frac{S_{\beta}(\omega)}{\omega^2\sqrt{1-t^2/\omega^2}} \text{d}\omega \,\, , \label{eq:cresonant} \\
& = \frac{\sigma^2_0 N_\beta}{ v} t^{1-\beta} g(\beta) h(\beta) \,\, , 
\label{eq:cfinal}
\end{align}
where $S_\beta(\omega)$ is given by Eq.~\eqref{eq:Sbeta} and $h(\beta)$ function is given by
\begin{align}
h(\beta) & \equiv \int_{1}^{\infty} \frac{\text{d}x}{x^{\beta+1}\sqrt{x^2 - 1}} = \int_{0}^{\pi/2} \cos^\beta(\alpha) \, \text{d}\alpha \,\, , \nonumber\\
& = \frac{\sqrt{\pi}}{2} \frac{\Gamma\left(\frac{\beta+1}{2}\right)}{\Gamma\left(\frac{\beta}{2}+1\right)} \,\, . \label{eq:h}
\end{align}
which is of the order of unity and decreases monotonically in the range of interest for $\beta$, as $h(1) \! =\! 1$ and $h(3/2)\! \approx \! 0.87$. In Eq.~\eqref{eq:cfinal} we see that for $\beta \! = \! 1$ the transition probability (i.e.~the error of the electron transfer process) is predicted to be {\it independent} of the tunnel coupling $t$. 

Equation \eqref{eq:cfinal} is the main analytical result of this paper: it shows  the scaling of the error probability of dot-to-dot transition probability with the noise power $\propto \sigma^2_0 N_\beta$, detuning sweep rate $v$, and tunnel coupling $t$. Such a scaling law can be shown in more straightforward way. The excitation here is caused by a transverse noise $\xi_\perp \sim 1/t \dot \xi(\tau)$, and hence their rate should be directly related to its noise spectrum at the energy gap between the ground and the excited state. As derived in in Appendix~\ref{app:OUderivative}, the spectrum of time-derivative of each TLF is given by $S_{\dot \xi}(\omega) = \omega^2 S_{\text{TLF}}(\omega)$.
 Using the model of $1/f^\beta$ noise described in Sec.~\ref{sec:1fmodel} we arrive at spectrum of derivative of such noise
	\begin{equation}
	S_{\beta,\dot{\xi}}(\omega) = \int \frac{N_\beta}{\gamma^\beta} \omega^2 S_\text{TLF}(\omega) \text{d}\gamma \sim |\omega|^{2-\beta}.
	\end{equation}
The probability of excitation is significant only near the anti-crossing, which corresponds to a time interval $\tau \in (-t/v,t/v)$ during which at least $|\bra{L}\ket{\vartheta(t/v)}|^2 \sim 0.15$ of the electron is delocalized between two dots. In these range of detunings, the energy gap can be approximated by its minimal value, the tunnel coupling $t$. As a result the excitation rate reads
\begin{equation}
\Gamma \propto \frac{1}{t^2}S_{1/\dot f^\beta}(\omega =t).
\end{equation}
From above one can reconstruct the probability of leaving the electron behind: it is given by a product of the above rate by time $\sim \! t/v$ that the state spends close to the anticrossing:
\begin{equation}
\langle |c_+^{(1)}(\infty)|^2 \rangle \propto \Gamma\, \frac{t}{v} = A\frac{\sigma^2}{v} t^{1-\beta}.
\end{equation}
where proportionality factor $A\sigma^2$ related to the noise amplitude is obtained by comparing the above qualitative formula with Eq.~(\ref{eq:cfinal}).

For $1/f$ noise characterized by power $S_{1}$ at frequency of $\omega_1 \! =\! 2\pi$ rad/s (i.e.~1 Hz), we have then
\begin{equation}
\langle |c_{+}(\infty)|^2\rangle_{\beta=1} = \frac{ S_{1} \omega_1}{2 v} \,\, .
\end{equation}
Let us notice now that typical order of magnitude of $\sqrt{S_{1}}$ is $1$ $\mu$eV/Hz$^{1/2}$ \cite{Freeman_APL16,Connors_arXiv19,Petit_PRL18,Mi_PRB18,Struck_arXiv19}, while maximal detuning sweep velocities allowing for fast and high-probability noiseless transfer for typical $t\approx 10$ $\mu$eV in Si QDs are of the order of 10-100 $\mu$eV/ns. Using these natural units for $S_{1}$ and $v$  we have 
\begin{equation}
\langle |c_{+}(\infty)|^2\rangle_{\beta=1} \approx 4.8 \frac{S_1 [\mu \text{eV}^{2}/\text{Hz}]}{v [\mu\text{eV}/\text{ns}]}  \,\, ,
\end{equation}
 which gives the error of about $5$\% for $S_{1} \! =\! 1$ $\mu$eV/Hz$^{1/2}$ and $v\! =\! 100$ $\mu$eV/ns. The realistic $1/f$ charge noise in detuning is thus expected to be very dangerous for high-fidelity electron transfer. 

\subsection{Results}
In this Section we describe the results of numerical simulations, in which we averaged the solutions of time-dependent Schr{\"o}dinger's equation in the four-dimensional subspace of interest over $N\! \sim \! 10^{4}$ realizations of $1/f^\beta$ noise. The latter was constructed from $1000$ sources of Ornstein-Uhlenbeck noise (individual fluctuators), with switching rates distributed according to distributions discussed in Sec.~\ref{sec:1fmodel}. These calculations will be compared with analytical formulas obtained in Sec.~\ref{sec:analytical}.

Recently reported values of $\sqrt{S_1}$ for charge noise affecting Si/SiGe quantum dots falls into range $\sqrt{S_1} = \sqrt{S(1 \,\,  \text{Hz})} \approx \! 0.2 - 2$ $\mu$eV/Hz$^{1/2}$ \cite{Freeman_APL16,Connors_arXiv19,Petit_PRL18,Mi_PRB18,Struck_arXiv19}, with $1/\omega$ behavior of the power spectral density in frequency range $\omega \! \sim \! t$.
As we have estimated in the previous Section, such noise is already sufficient to significantly limit the adiabatic transition probability at low and moderate sweep speed. On the other hand, the noiseless Landau-Zener process of nonadiabtic transition becomes relevant at high sweep speed. Consequently, in presence of charge noise we expect a finite window of sweep rates at which the transfer probability is larger than a certain value.  

\begin{figure}[tb]
	\centering
	\includegraphics[width = \columnwidth]{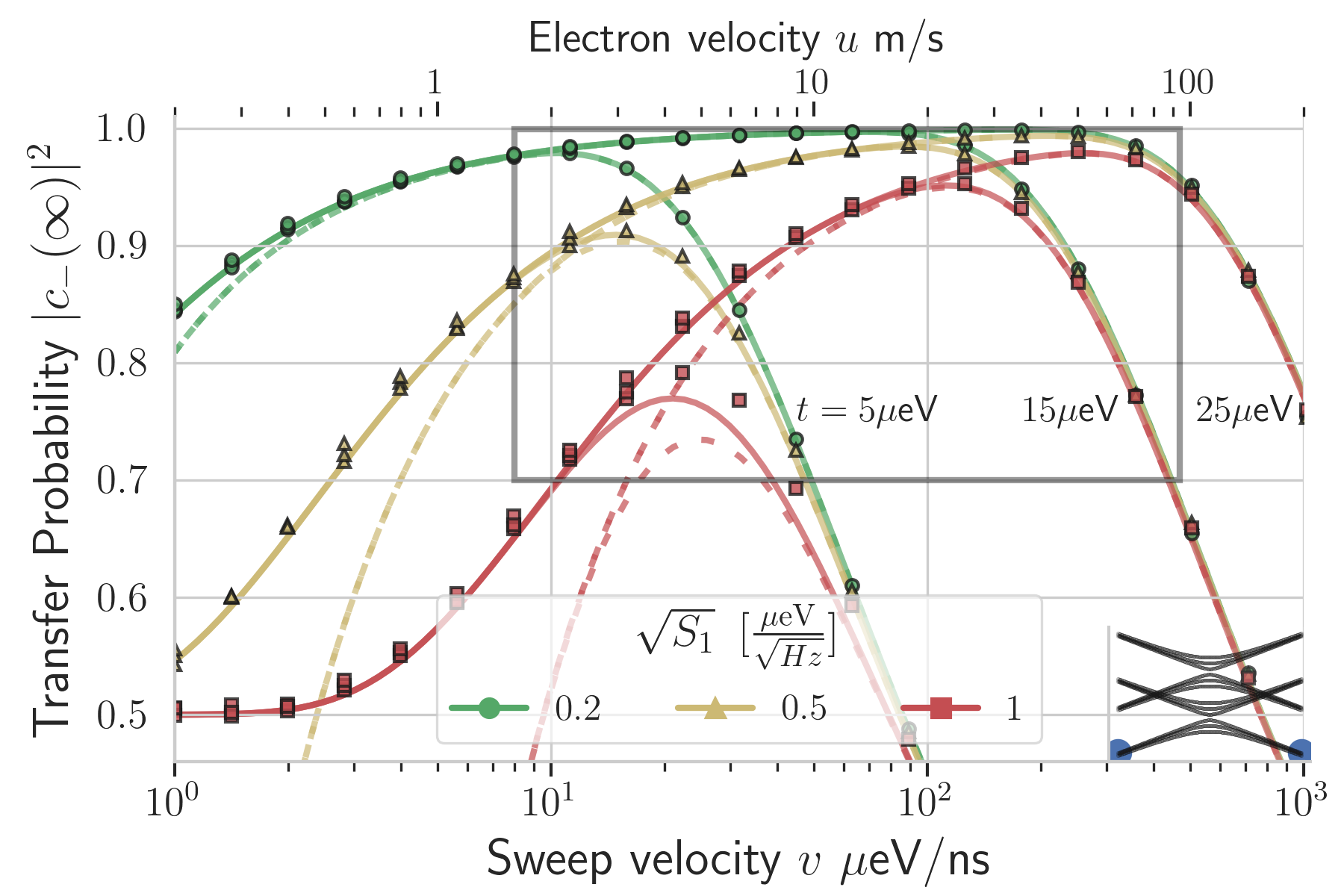}
	\includegraphics[width = \columnwidth]
	{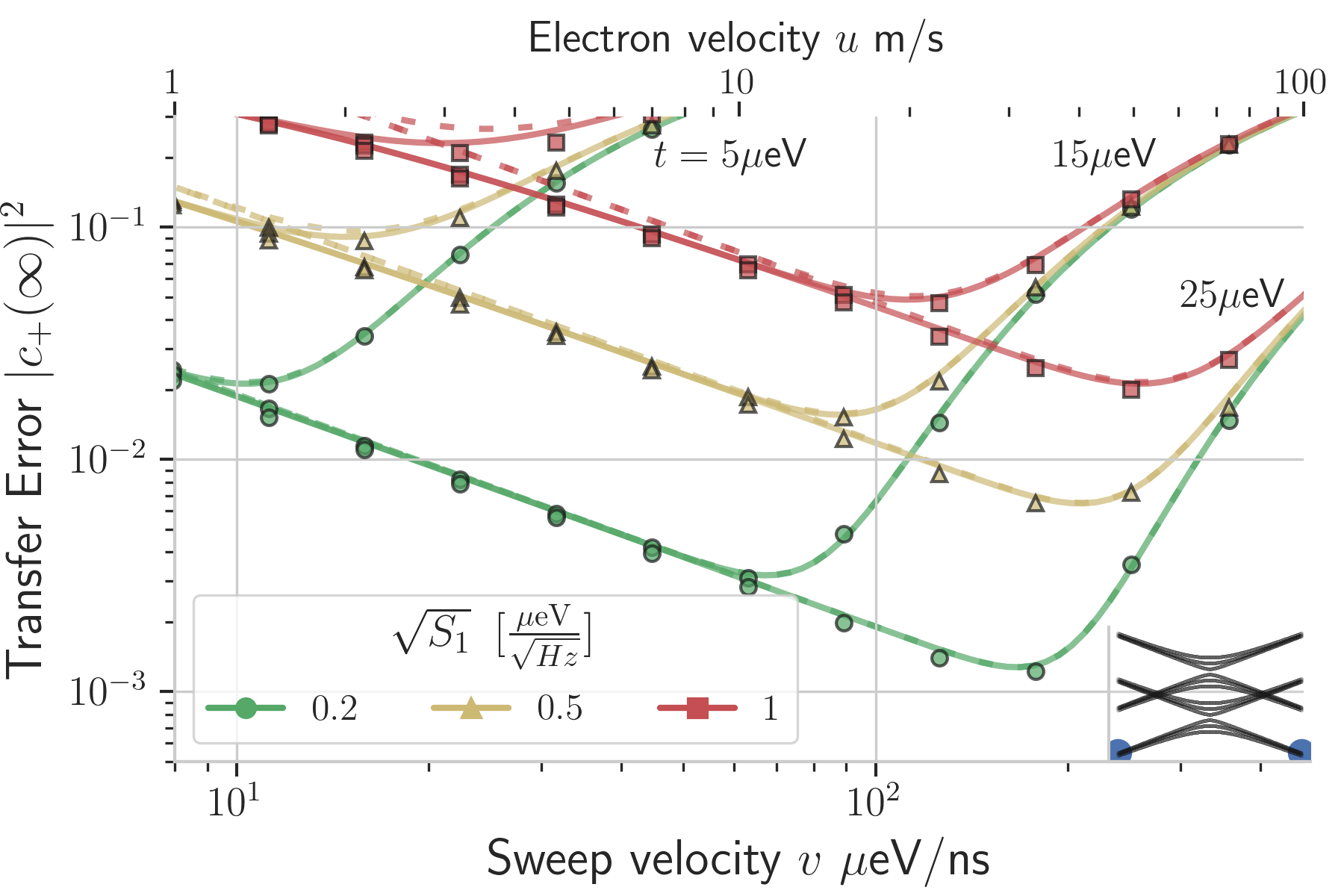}
	\caption{Top: Probability of successful electron transfer $p_R\! =\! 1-p_L$ (i.e. electron ending in right dot); Bottom: Probability of non-adiabatic excitation, the error $p_L$ (i.e. the probability of leaving the electron in the left dot). Results of simulations for $1/f$ noise in detuning characterized by $\sqrt{S(1 \, \mathrm{Hz})} \! =\! 0.2$, $0.5$, and $1$ $\mu$eV/Hz$^{1/2}$ are marked by symbols, which are compared with Eq.~(\ref{eq:cfinal}), shown as dashed line, and a re-exponentiation of this result (see text), shown as solid lines, which agrees very well with numerical simulations in the whole range of results. Bottom panel corresponds to the region marked by a black rectangle on top. The results for tunneling $t \! =\! 5$, $15$, and $25$ $\mu$eV converge at low $v$, when the value of $p_{R} \! < \! 1$ is determined only by noise, and it is independent of $t$. For larger $v$, the transition to Landau-Zener dependence due to deterministic nonadiabatic effects described by Eq.~(\ref{eq:PLZ}) occurs at $v\! \propto \! t^2$. For considered noise power the error always exceeds $10^{-3}$.}
	\label{fig:1f1}
\end{figure}

This situation in case of $\beta\! =\! 1$ and transfer of an electron initialized in the lowest energy state, is depicted in Fig.~\ref{fig:1f1}, where numerical simulations (points) are compared with predictions of Eq.~(\ref{eq:cfinal}) plotted as dashed lines.
Solid lines correspond to result in which we re-exponentiate the short-time behavior of transfer probability, i.e.~we replace a difference in occupation $|c_{-}(\infty)|^2-|c_{+}(\infty)|^2$ by $\text{exp}(-2|c_{+}^{(1)}(\infty)|^2)$, and $|c_+(\infty)|^2 = \frac{1}{2}\big(1 + e^{-2|c_+^{(1)}|^2}\big)$ as a consequence.
In noise dominated regime (left part of the figure), decay of probability is insensitive the value of tunnel coupling $t$, which follows from scaling of transfer error with $t^{\beta - 1}$ in Eq.~(\ref{eq:cfinal}). For the highest considered amplitude of noise, corresponding to $\sqrt{S_1} = 1\mu$eV/$\sqrt{\text{Hz}}$, as $v$ decreases below $\sim \! 10 $ $\mu$eV/ns, the probability of successful transfer tends to $1/2$, which means that during time the electron spends in the energy level anticrossing region (i.e.~when the electron is delocalized between the two dots), the influence of classical (i.e.~effectively infinite temperature) noise is strong enough to equalize the occupations of the two adiabatic energy levels. 
Higher values of $t$ of course allow for increasing $v$ to larger values without entering the regime in which the error is dominated by the Landau-Zener process.
For instance, for $\sqrt{S_1} = 0.5\mu$eV/$\sqrt{\text{Hz}}$ measured recently in Si/SiGe quantum dot \cite{Struck_arXiv19}, achievable transfer probability may vary from  $90$\% to $99\%$ monotonically depending on the tunnel coupling $t = 5 - 25 \mu$eV.   

\begin{figure}[tb]
	\centering
	\includegraphics[width = \columnwidth]{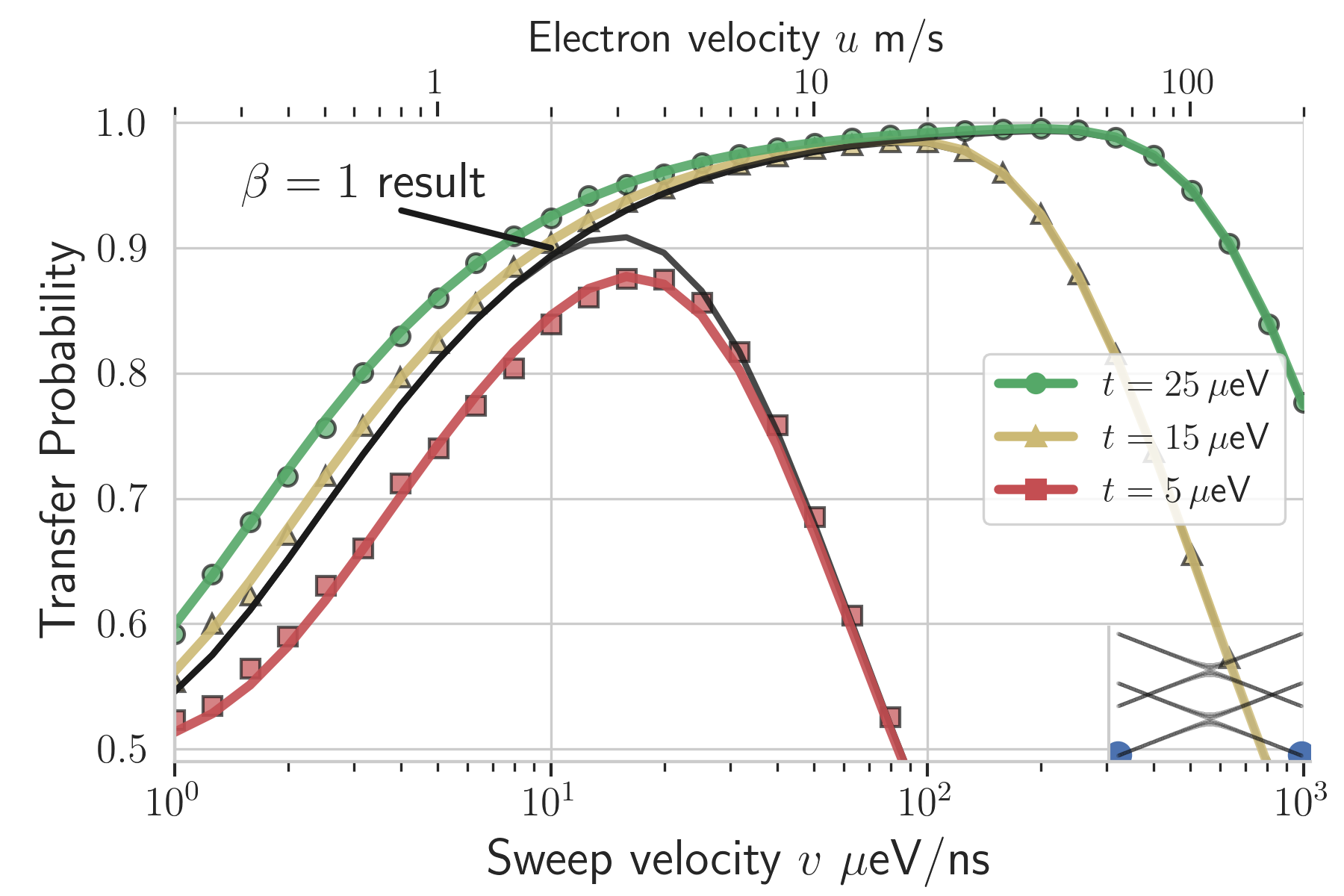}
	\caption{Simulated probability of nonadiabatic excitation (i.e.~leaving the electron behind in the left dot) for $1/f^{\beta}$ noise with $\beta\! =\! 1.5$, characterized by $\sqrt{S(1 \, \mathrm{Hz})} \! =\! 0.5\,\mu$eV/Hz$^{1/2}$. The simulated results (symbols) for tunneling $t \! =\! 5$, $15$, and $25$ $\mu$eV do not converge to the same dependence at low $v$ and fit the exponential curve (sold lines).
	}
	\label{fig:1f15}
\end{figure}

In Fig.~\ref{fig:1f15} we show analogous results for the case of $\beta\! =\! 1.5$, i.e.~noise with relatively larger low-frequency component. In order to make a meaningful comparison to the $\beta\! =\! 1$ case, we chose the noise amplitudes in such a way, that they agree at frequency $\omega_h$  corresponding to tunnel coupling of $15$ $\mu$eV. We have then the spectral density of noise with $\beta\! \neq \! 1$
$$  S_\beta(\omega) =  \nonumber\frac{S_1 \omega_1}{\omega_h} \left(\frac{\omega_h}{\omega}\right)^\beta \,\, ,$$
in which $S_1$ are the previously used parameters characterizing the power of $1/f$ noise. In Fig.~\ref{fig:1f15} we show results for $\sqrt{S_1}\! =\! 0.5$ $\mu$eV/Hz$^{1/2}$ and three values of tunnel coupling $t$.
Contrary to the previously discussed case of $\beta = 1$ (shown in Fig.~\ref{fig:1f15} as black solid lines corresponding to three values of $t$), now the probability of transfer diminishes with decreasing $t$ also on the low-speed side of the plot, according to scaling $\propto t^{1 - \beta} = t^{-0.5}$.

Let us note now that if we focus on the regime of $v$ lower than the optimal one (the left side of the plot), transfer probability is increased relatively to $\beta\! =\! 1$ result when we consider noise with stronger low-frequency component. For cleanest comparison let us focus on results for both values of $\beta$ for $t\! =\! 15$ $\mu$eV, corresponding to frequency $\omega_h$ at which both spectra take on the same value. The ratio of numerically calculated probabilities is in fact well approximated by $h(1.5)/h(1) \approx 0.88$. This follows from the fact that when we plug in the parameters of $1/f^{\beta}$ noise considered here into Eq.~\eqref{eq:cfinal} we obtain 
	\begin{equation}
	\langle |c_{+}(\infty)|^2 \rangle_{\beta\neq 1} 	= \langle |c_{+}(\infty)|^2 \rangle_{\beta = 1} \left(\frac{t}{ \omega_h}\right)^{1-\beta} h(\beta) \,\, ,
	\end{equation}
	and $h(\beta)$ is a monotonically decreasing function of $\beta$.

\begin{figure}[tb]
	\centering
	\includegraphics[width = \columnwidth]{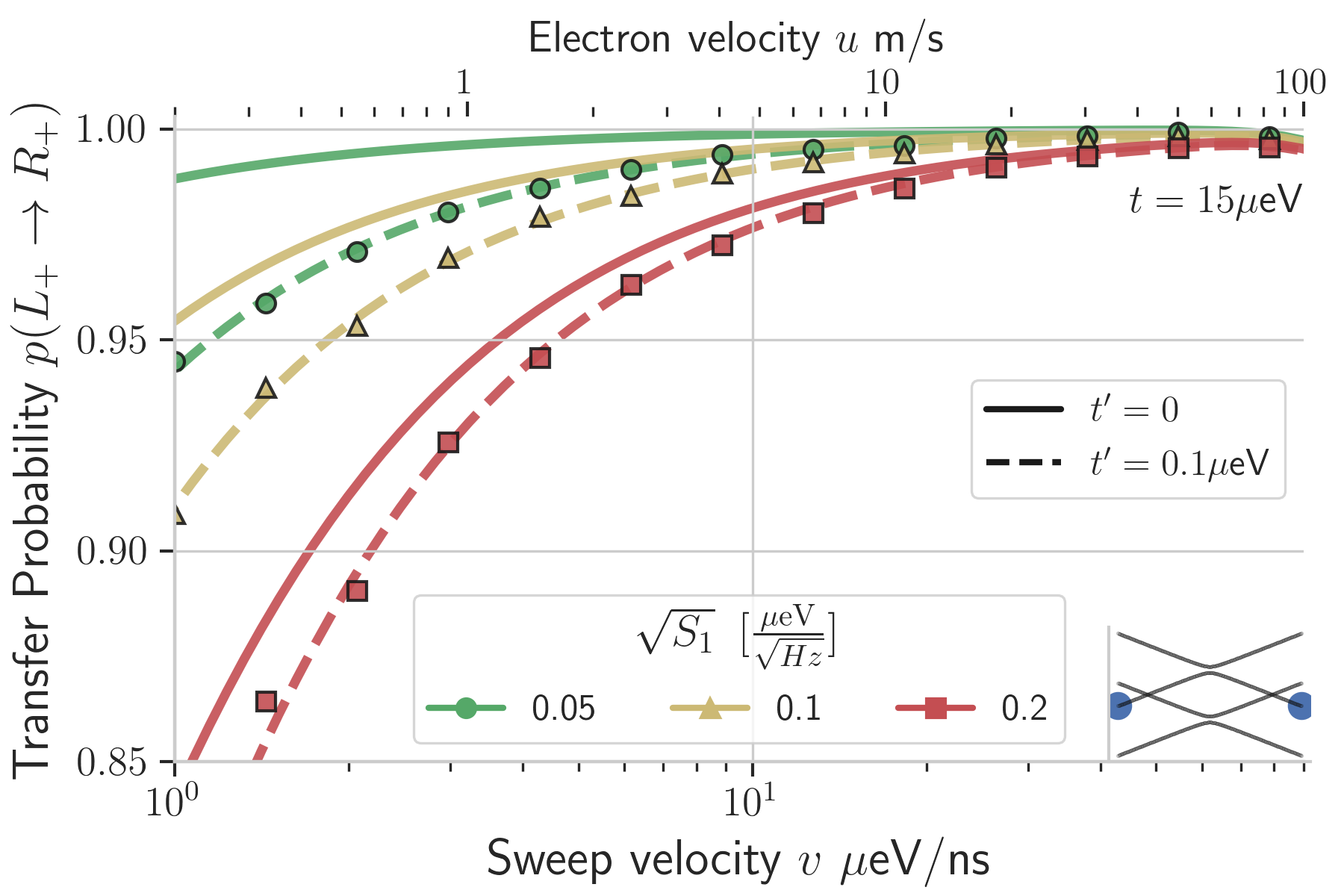}
	\caption{Probability of successful electron transfer calculated for electron initialized in higher energy state (spin up) for $t' \! =\! 0$, $0.1$ $\mu$eV and $1/f$ noise (with parameters as in Fig.~\ref{fig:1f1}). Symbols correspond to numerical simulation of influence of noise for $t'\! =\! 0.1$ $\mu$eV. Solid lines are analytical result for $t'\!=\! 0$, which are the same as results for the ground state shown in Fig.~\ref{fig:1f1}). Dashed line is a product of Eqs.~\eqref{eq:cl} and $1-|c_{-}(\infty)|^2$ where $|c_{-}(\infty)|^2$ is given by Eq.~\eqref{eq:cfinal}, i.e.~it corresponds to analytical formula accounting approximately for all the effects of noise in presence of $t'\! \neq \! 0$.}
	\label{fig:1f1E}
\end{figure}

Finally, let us analyze the effect of dynamic noise on the transfer of the electron initialized in the excited state. As we have discussed in Section \ref{sec:nonoise}, the transfer in this case differs from the previously considered case only when additional anticrossings, characterized by $t'\!\neq \! 0$, are affecting the evolution of this state. As shown in Fig.~\ref{fig:qsnoise}, for $t'\! =\! 0.1$ $\mu$eV, measured recently in Si/SiO$_{2}$ quantum dots \cite{Harvey-Collard_PRL19,Tanttu_PRX19}, the probability of transfer is lowered to $\approx \! 0.95$ at $v\! \approx \! 1$ $\mu$eV/ns when the oscillations caused by interference of two transfer paths are dephased. While in Sec.~\ref{sec:nonoise} we have discussed quasi-static Overhauser field noise as the source of this dephasing, one should expect an analogous effect to be caused by the slowest dynamic fluctuations of detuning that induce even small random changes in the electron transfer time. In Fig.~\ref{fig:1f1E} we show that this effect is indeed present for $1/f$ noise, as the numerically calculated transfer probabilities do not exhibit any oscillations for $t'\! =\! 0.1$ $\mu$eV. There are now two effects of noise on transfer: the previously discussed excitation to the other adiabatic state by high-frequency noise, and the dephasing of interference of state propagation along two paths (due to $t'\! \neq \! 0$) due to lower-frequency noise. The latter effect is irrelevant at higher $v$, for which the results for $t'\! \neq \! 0$ and $t'\! =\! 0$ are indistinguishable (and of course the same as the result for ground state transfer). Due to this in Fig.~\ref{fig:1f1E} we focus on the low $v$ range. The numerical simulations for $t' \! =\! 0.1$ $\mu$eV are shown as symbols, solid lines are $t'\! =\! 0$ results, and the dashed lines represent the product of Eqs.~\eqref{eq:cl} and $1-|c_{-}(\infty)|^2$ where $|c_{-}(\infty)|^2$ is given by Eq.~\eqref{eq:cfinal}.
The main result is that for the considered value of $t'$, transfer error at low $v$ is limited by path interference and its dephasing due to slow noise for noise characterized by $S_{1} \! \leq \! 0.1$ $\mu$eV/Hz$^{1/2}$. 

\begin{table}[tb!]
	\begin{tabular}{c||c|c|c}
		Ground state & \multicolumn{3}{c}{Tunneling $t\,\, [\mu$eV$]$}           \\ \cline{1-4} 
		$\sqrt{S(1\text{Hz})}\, \left[\tfrac{\mu \text{eV}}{\sqrt{\text{Hz}}}\right]$ & $5$            & $15$            & $25$            \\ \hline 
		$0.2$                  & $0.978\, @ \, 10.9$ & $0.997 \, @ \, 67.2$  & $0.999 \, @ \,174.2$ \\ \hline
		$0.5$                  & $0.905\, @ \, 14.6$ & $0.983 \, @ \, 85.3$ & $0.993 \, @ \,206.6$ \\ \hline
		$1.0$                  & $0.780\, @ \, 21.9$  & $0.950 \, @ \, 106.4$ & $0.978 \, @ \, 237.0$
	\end{tabular}\qquad 	\begin{tabular}{c||c|c|c}
	Excited state& \multicolumn{3}{c}{Tunneling $t\,\, [\mu$eV$]$}           \\ \cline{1-4} 
	$\sqrt{S(1\text{Hz})}\, \left[\tfrac{\mu \text{eV}}{\sqrt{\text{Hz}}}\right]$ & $5$            & $15$            & $25$            \\ \hline 
	$0.2$                  & $0.973\, @ \, 10.3$ & $0.996 \, @ \, 67.2$  & $0.999 \, @ \,174.2$ \\ \hline
	$0.5$                  & $0.904\, @ \, 14.5$ & $0.983 \, @ \, 85.3$ & $0.993 \, @ \,196.85$ \\ \hline
	$1.0$                  & $0.780\, @ \, 20.1$  & $0.950 \, @ \, 113.9$ & $0.978 \, @ \, 250.0$
\end{tabular}
	\caption{ Maximal probability of ground (upper table) and excited (lower table) state transfer with corresponding optimum of sweep velocity $p_{\max} \,@ \, v_{\text{opt}} [\mu\text{eV}/ns]$ for different value of tunneling $t$ and noise power at $1$Hz for $S(\omega)\propto 1/\omega$. We have used  $t' \!= \! 0.1 \mu$eV which is of the order of maximal value observed in SiMOS devices.} \label{tab:optimal}
\end{table}

The minimal probabilities of successful electron transfer and corresponding optimal velocities for all the sets of parameters used in all the Figures are gathered in Table \ref{tab:optimal}. We see that in order for error rate to be lower than $1$\%, the tunneling must be at least $25$ $\mu$eV, or $\sqrt{S_{1}}$ should be below $0.5$ $\mu$eV/Hz$^{1/2}$.

\section{Discussion and conclusion}
We have considered the influence of $1/f^\beta$ noise in detuning on the process of adiabatic electron charge transfer between two quantum dots. The probability of noise-induced excitation that leads to the electron being left behind is proportional to $\sigma^2 t^{1-\beta}/v$, where $\sigma^2$ characterizes the total noise power, $t$ is the tunnel coupling, and $v$ is the detuning sweep speed. 
Our main result is that for a realistic set of parameters - $1/f$ noise with amplitude of $\sim \! 1$ $\mu$eV/Hz$^{1/2}$ at $f\! =\! 1$ Hz, $t\! \approx \! 10$ $\mu$eV, $v\! \sim 10-200$ $\mu$eV/ns corresponding to real space speed of $2-40$ m/s assuming interdot distance of $50$nm  and maximum detuning swing of $250 \mu$eV -  this probability is at best $\sim 1$ \%. Consequently, scaling up the ``electron conveyer belts'' to systems of $N\! \sim \! 100$ quantum dots from recently demonstrated $N\! =\! 9$ case \cite{Mills_NC19} will require lowering of the charge noise amplitude, or using higher tunnel couplings, for which the transfer can be done more quickly without triggering the noiseless nonadiabatic Landau-Zener excitation process. 

The fact that the charge transfer error probability scales with $t^{1-\beta}$ has an additional practical consequence. For a many-dot chain with inhomogeneous values of $t$, and for $1/f$ charge noise ($\beta=1$), assuming that the rms amplitudes of charge noise on each dot are similar, the noise-induced errors in charge transfer are independent of values of $t$, so when considering the {\it upper} limit on $v$ that can be tolerated given a spread of values of $t$, one only needs to consider the Landau-Zener formula for probability of nonadiabatic transition in the noiseless case, i.e.~$p_L\! =\! \text{exp}(-\pi t^{2}/2v)$. The lower limit on $v$ is on the other hand determined by maximal noise amplitude encountered along the chain of dots.

We have also analyzed the noiseless situation for the case of transfer of an electron in an excited state (higher-energy spin state), for which there are two additional energy level anticrossings due to $t' \! \ll \! t$ tunnel couplings corresponding to tunneling with a spin flip due allowed by spin orbit interaction. For relatively fast electron transfer the electron should traverse these anticrossing non-adiabatically (so that they are in fact effectively level crossings)
However, when $t'$ is $\approx 0.01 t$, the sweep rate is bounded from below by finite probability of adiabatic transition through $t'$ anticrossings. Below this rate, the state can go through two paths, and the interference of these processes leads to very rapid (as function of $v$) oscillations of transfer probability. In presence of quasi-static noise (due to fluctuations of Overhauser splittings in the two dots), or the slowest components of $1/f$ noise that still lead to finite fluctuation of detuning during the transfer process, these oscillations become dephased, but the transfer probability is lowered compared to $t'\! =\! 0$ case. This could make the window of $v$ allowing for high-fidelity electron transfer quite narrow in SiMOS devices, where such values of $t'$ were measured \cite{Harvey-Collard_PRL19,Tanttu_PRX19}, unless one chooses the magnetic field direction and geometry of the structure that makes $t'$ smaller.

\section*{Acknowledgments}
We would like to thank Lars Schreiber for discussions. 
This work has been funded by the National Science Centre (NCN), Poland under QuantERA programme, Grant no.~2017/25/Z/ST3/03044. 
This project has received funding from the QuantERA Programme under the acronym Si QuBus. 

\appendix

\begin{widetext}
	\section{Correlation function of time derivative of Ornstein-Uhlenbeck process} \label{app:OUderivative}
	We calculate time integral involving correlation function of derivative of Ornstein-Uhlenbeck process fulfilling Langevin equation $\dot \xi(\tau) = -\frac{1}{\tau_c} \xi(\tau) + \sqrt{\frac{2\sigma^2}{\tau_c}} w(t)$, where $w(t)$ is derivative of the Wiener process $w(\tau)=dW_\tau/d\tau$, known as the white noise with correlation function $\langle w(\tau_1)w(\tau_2) \rangle = \delta(\tau_1-\tau_2)$. We would be particularly interested in the integral of a form: 
	\begin{align}
	&\int_{-T}^{T} \text{d}\tau_1 \text{d}\tau_2 f(\tau_1) f^*(\tau_2) \langle \dot\xi(\tau_1)\dot\xi(\tau_2) \rangle  = \int_{-T}^{T} \text{d}\tau_1 \int_{-T}^{T} \text{d}\tau_2  \frac{2\sigma^2}{\tau_c} \langle w(\tau_1) w(\tau_2) \rangle  f(\tau_1) f^*(\tau_2) 	 \\ &\quad+2 \int_{-T}^{T} \text{d}\tau_1 \int_{-T}^{\tau_1} \text{d}\tau_2\bigg(\frac{\langle \xi(\tau_1) \xi(\tau_2) \rangle}{\tau_c^2} 
	- \frac{\sqrt{2 \sigma^2}}{\tau_c^{3/2}}\big(\langle \xi(\tau_1)w(\tau_2)\rangle + \underbrace{\langle\xi(\tau_2)w(\tau_1) \rangle}_0\big)\bigg) \bigg( f_1(\tau_1) f_2^*(\tau_2) + f_1^*(\tau_1) f_2(\tau_2)\bigg). \nonumber
	\end{align}
One can use solution to the Langevin equation $\xi(\tau) = \sqrt{\frac{2\sigma^2}{ \tau_c}}\int_{-T}^{\tau} \text{d}s e^{-\gamma(\tau-s)} w(s)$ to simplify the second term, which can be written as:
	\begin{align}
	\int_{-T}^{T} \text{d}\tau_1 \int_{-T}^{\tau_1} \text{d}\tau_2 &\bigg(\frac{\sigma^2}{\tau_c^2} e^{-\frac{\tau_1-\tau_2}{\tau_c}} - \frac{\sqrt{2 \sigma^2}}{\tau_c^{3/2}}\, \sqrt{\frac{2\sigma^2}{\tau_c}}\, \int_{-T}^{\tau_1}\text{d}s  e^{-\frac{\tau_1-s}{\tau_c}} \langle w(s)w(\tau_2) \rangle\bigg) \bigg( f(\tau_1) f^*(\tau_2) + f^*(\tau_1) f(\tau_2)\bigg) \nonumber \\ 
	&= \int_{-T}^{T} \text{d}\tau_1 \int_{-T}^{\tau_1} \text{d}\tau_2 \bigg(\frac{\sigma^2}{\tau_c^2} e^{-\frac{\tau_1-\tau_2}{\tau_c}} - \frac{\sqrt{2 \sigma^2}}{\tau_c^{3/2}}\, \sqrt{\frac{2\sigma^2}{\tau_c}}\,   e^{-\frac{\tau_1-\tau_2}{\tau_c}}\bigg) \bigg( f(\tau_1) f^*(\tau_2) + f^*(\tau_1) f(\tau_2)\bigg)
	\end{align}
Altogether this allows to write
	\begin{equation}
	\int_{-T}^{T}\int_{-T}^{T} \text{d}\tau_1 \text{d}\tau_2 \langle \dot\xi(\tau_1)\dot\xi(\tau_2) \rangle f(\tau_1) f^*(\tau_2) = \frac{2\sigma^2}{\tau_c} \int_{-T}^{T} \text{d}\tau_1\int_{-T}^{T} \text{d}\tau_2 f(\tau_1)f^*(\tau_2)  \delta(\tau_1-\tau_2)  - \frac{1}{2\tau_c}e^{-|\tau_1-\tau_2|/\tau_c} \bigg)
	\end{equation} 
In the Appendix \ref{app:OUcorrection} we will use derived result for $f(\tau) = \frac{t}{r_0^2(\tau)} \exp{- i\int_{-T}^\tau r(\tau')}$.
	
	\section{Transition probability due to Ornstein-Uhlenbeck noise} \label{app:OUcorrection}
We calculate here the integral in Eq.~\eqref{eq:first_ord},
\begin{equation}
\langle |c_+^{(1)}(\infty)|^2 \rangle   = \frac{t^2}{4} \int_{-\infty}^{\infty}\text{d} \tau_1  \text{d}\tau_2 \frac{\langle \dot \xi(\tau_1) \dot \xi(\tau_2) \rangle }{\big[r_0(\tau_1)r_0(\tau_2)\big]^2}\exp{-i\int_{\tau_2}^{\tau_1} \text{d}\tau' \,r_0(\tau')}  \label{eq:c1}
\end{equation}
for an Ornstein-Uhlenbeck process, for which $\langle \dot \xi(\tau_1) \dot \xi(\tau_2)  \rangle = \frac{\sigma^2_0}{\tau_c} \left(2\delta (\tau_1-\tau_2) - \tfrac{1}{\tau_c}e^{-|\tau_1-\tau_2|/\tau_c}\right) $ as derived in Appendix \ref{app:OUderivative}. In Fourier space it can be written in terms of spectrum of the noise derivative,
	\begin{equation}
	2\delta (\tau_1-\tau_2) - \tfrac{1}{\tau_c}e^{-|\tau_1-\tau_2|/\tau_c} = 2 \int_{-\infty}^{\infty}\frac{\text{d}\omega}{2\pi} \left( e^{i\omega(\tau_1-\tau_2)} - \frac{1}{1 + \omega^2 \tau_c^2} e^{i\omega(\tau_1-\tau_2)} \right) = \int_{-\infty}^{\infty} \frac{2\omega^2 \tau_c^2}{1 + \omega^2 \tau_c^2} e^{i\omega(\tau_1 - \tau_2)},
	\end{equation}
and then substituted into \eqref{eq:c1},
	\begin{equation}
	\frac{\sigma^2_0 t^2}{4 \tau_c} \int_{-\infty}^{\infty} \frac{\text{d}\omega}{2\pi} \frac{2\omega^2 \tau_c^2}{1 + \omega^2\tau_c^2}  \left(\int_{-\infty}^{\infty} \text{d} \tau_1 \frac{1}{r_0^2(\tau_1)} \exp{i\omega \tau_1 - i\int_{0}^{\tau_1} r_0(\tau')} \right)  \left(\int_{-\infty}^{\infty} \text{d}\tau_2  \frac{1}{r_0^2(\tau_2)} \exp{-i\omega \tau_2 + i\int_{0}^{\tau_2} r_0(\tau')}\right)
	\end{equation}
	We calculate time integrals using stationary phase method, within which in the leading order we have $\int g(\tau) e^{ih(\tau)} \text{d} \tau \approx \sum_{\tilde \tau} g(\tilde \tau) e^{if(\tilde \tau)} \int e^{i f''(\tilde \tau)\frac{(x-\tilde \tau)^2}{2}} \text{d}x$, where $\tilde \tau$ denotes stationary time, which fulfills equation $f'(x_0) = 0$. Here $\tilde \tau$ is found from equation $\partial_\tau (\omega \tau - \int_{0}^{\tau} r_0(\tau')) = 0 \implies \omega = r_0(\widetilde \tau)$ and reads $\tilde \tau = \pm \frac{\sqrt{\omega^2 - t^2}}{v}$, while $f''(\pm\widetilde \tau) = \partial^2_\tau (\omega \tau - \int_{0}^{\tau} r_0(\tau'))|_{\tau = \pm\widetilde \tau} = \mp\frac{v}{\omega} \sqrt{\omega^2 - t^2}$. As a result we have:
	\begin{align}
	\label{int_phase}
	\int_{-\infty}^{\infty} &\text{d} \tau_1 \frac{e^{i\omega\tau_1 - i\varphi(\tau_1)}}{r_0^2(\tau_1)} \approx \frac{e^{i\omega\widetilde\tau - i \varphi(\widetilde \tau))}}{r_0^2(\widetilde\tau)}\int_{-\infty}^{\infty} \text{d}\tau_1\, e^{ig''(\tilde \tau)\frac{(\tau_1 - \widetilde \tau)^2}{2}}+ \bigg( \widetilde\tau \to -\widetilde \tau \bigg) =  
	\sqrt{\frac{8\pi}{v\omega^3\sqrt{\omega^2-t^2}}} \cos(\omega\widetilde\tau - \varphi(\widetilde \tau) - \frac{\pi}{4})
	\end{align}
	We denoted dynamical phase as $\varphi(\widetilde \tau) = \int_{-\infty}^{\widetilde \tau} r_0(\tau') \text{d}\tau'$. Since second integral is the same, as a complex conjugate of the first, we write
	\begin{align}
	\frac{4 \pi \sigma^2_0 t^2}{v \tau_c} \int_{t}^{\infty} \frac{\text{d}\omega}{2\pi} \frac{2\omega^2 \tau_c^2}{1 + \omega^2\tau_c^2} \frac{1}{\omega^4\sqrt{1 - \frac{t^2}{\omega^2}}} \cos^2\left(\omega \widetilde \tau - \varphi(\widetilde \tau) - \frac{\pi}{4}\right) = \frac{\sigma^2_0 t^2}{ v \tau_c} \int_{t}^{\infty} \text{d}\omega \frac{2\omega^2 \tau_c^2}{1 + \omega^2\tau_c^2} \frac{1 + \sin(2\omega \widetilde \tau - 2\varphi(\widetilde \tau))}{\omega^4\sqrt{1 - \frac{t^2}{\omega^2}}}. 
	\end{align}
	Now quickly oscillating term $\propto \sin(2\omega \tilde \tau - 2 \int r_0(\tau'))$ can be neglected which leads to final result:
	\begin{equation}
\langle |c_+^{(1)}(\infty)|^2 \rangle \approx \frac{\sigma^2_0 t^2}{2v \tau_c} \int_{t}^{\infty}  \frac{\text{d}\omega}{\omega^4\sqrt{1 - \frac{t^2}{\omega^2}}}\frac{2\omega^2 \tau_c^2}{1 + \omega^2\tau_c^2}  = \frac{ t^2}{2v} \int_{t}^{\infty} \frac{\text{d}\omega}{\omega^2\sqrt{1 - \frac{t^2}{\omega^2}}}S_\text{ou}(\omega) = \frac{\sigma^2_0 }{v} R(t \tau_c)  \,\, .
	\end{equation}  
In the above we have defined function
\begin{equation}
R(t\tau_c) = (t\tau_c)^2 \int_{t\tau_c}^{\infty} \frac{\text{d} x}{x^2 + x^4} \left(1 - \frac{t^2\tau_c^2}{x^2}\right)^{-\frac{1}{2}} = \frac{\pi}{2}t\tau_c\left(1 - \frac{1}{\sqrt{1+\frac{1}{t^2\tau_c^2}}}\right)
\end{equation}

\end{widetext}	


%

\end{document}